\documentstyle[epsf,12pt,a4]{article}
\newcommand{\be}{\begin{equation}}
\newcommand{\ba}{\begin{eqnarray}}
\newcommand{\ee}{\end{equation}}
\newcommand{\ea}{\end{eqnarray}}
\newcommand{\nn}{\nonumber}

\newcommand{\GeV}{\;\mbox{GeV}}
\newcommand{\MeV}{\;\mbox{MeV}}
\newcommand{\eV}{\;\mbox{eV}}
\newcommand{\keV}{\;\mbox{keV}}

\newcommand{\secs}{\;\mbox{s}}

\newcommand{\simgt}{\stackrel{>}{{}_\sim}}
\newcommand{\ol}{\overline}

\newcounter{currequation}
\begin{document}
\title{
Relating a small decrease of $(m_p/m_e)$ with cosmological time to a small
cosmological constant
}
\author{Saul Barshay\footnote{barshay@kreyerhoff.de} and Georg Kreyerhoff\footnote{georg@kreyerhoff.de}\\
III.~Physikalisches Institut A\\
RWTH Aachen\\
D-52056 Aachen}
\maketitle
\begin{abstract}
The possible small variation downward by about $10^{-5}$, of the ratio of the proton mass
to the electron mass, over  cosmological time
is related to the decrease with time of a small vacuum expectation value for a Goldstone-like
pseudoscalar field. 
The initial vacuum expectation value controls the magnitude of a very small cosmological constant.
\end{abstract}
Recently, new data \cite{ref1} has given an indication that the ratio of the proton to electron
mass $\mu = (m_p/m_e)$, has decreased over a cosmological time interval. If interpreted in terms of
an effective decrease in the proton mass, the data suggest a decrease by about $(10 \eV)\times \mu
\sim 18 \keV$, over a period of about the past twelve billion years. Natural questions which arise
are then the following.
\begin{itemize}
\item[(1)] Can one obtain the direction of change, a decrease, independently
of the possibility that coupling parameters such as $\alpha_{\mathrm{em}}$, depend upon time? Data
on  the latter possibility \cite{ref2,ref3}, have stimulated the search for time variation of
physical ``constants'' \cite{ref4,ref1}. Recent data \cite{ref5,ref6} have not yet confirmed a
variation \cite{ref2,ref3}. 
\item[(2)] Can one obtain an estimate of the small absolute scale of the mass decrease, a few
keV, by relating it to other small energy-scale effects, like
a cosmological constant of about $3\times 10^{-47} \GeV^4$, and possibly to
a neutrino mass of less than 0.1 eV?
\item[(3)] Is it possible to estimate the cosmological time scale $\overline{t}$, for significant
mass decrease (on the above small energy scale), in terms of another cosmological parameter 
associated with the early universe?
\end{itemize}
The purpose of this note is to illustrate how  there
might be affirmative answers to these questions. 
The basis is the standard
assumption that the main contributions to particle mass arise from the nonzero
vacuum expectation values of spin-zero fields. It is usually assumed that such a
vacuum expectation value arises at the stable minimum of some effective potential-energy
density. Here, the first assumption is that a small, particular
vacuum expectation value of a pseudoscalar, Goldstone-like field $b$ (which spontaneously
breaks CP invariance in a cosmological context \footnote{
Spontaneous $CP$ violation is a
motivation for considering a nonzero vacuum expectation value for the $b$ field.
For a brief time near to
$\sim 10^{-36}\secs$, there can be a
CP-violating effect such as an antineutrino-neutrino asymmetry from a primary,
radiation-producing decay process \cite{ref7,ref11}. 
}) is a 
metastable maximum value (i.~e.~over
a cosmological time interval $\overline{t}$) \cite{ref7}.
We note that the possibility that the standard, scalar-field 
inflationary dynamics in the very
early universe originates at a maximum of an effective potential for a (classical)
scalar (inflaton) field $\phi$, has already been considered in detail \cite{ref8,ref9,ref10}.
It has been shown that radiative corrections to a $\lambda \phi^4$ potential-energy density
can set up an effective energy density with a maximum, with $\phi$  at or just above the Planck mass
$M_{\mathrm{Pl}} \cong 1.2 \times 10^{19}\GeV$, {\underline{and a minimum}} with $\phi$ just below
$M_{\mathrm{Pl}}$ \cite{ref9}. Inflation occurs 
with the inflaton field at the maximum and during its movement
to the minimum. The positive second derivative of the effective potential with respect
to the field variable, at the minimum, corresponds to the large squared mass of inflaton
quanta (estimated \cite{ref7,ref8} to be $m_\phi^2 \sim (5\times 10^{11}\GeV)^2$). If metastable
(i.~e.~essentially decoupled from big-bang radiation) these quanta can constitute dark
matter today \cite{ref7,ref8,ref11}. The pseudoscalar field $b$ can be connected with the
scalar $\phi$, in a hypothetical, idealized model of a cosmological, spontaneously-broken chiral
symmetry \cite{ref7,ref8}. However, the pseudoscalar field $b$ is a separate hypothesis from the
scalar field $\phi$ whose vacuum energy density generates the hypothetical inflation near $t=0$,
and the $b$-field dynamics over cosmological time intervals is distinct.
The hypothetical, small vacuum expectation value $b_0$ of the $b$ field,
contributes a vacuum energy density  $\lambda b_0^4
\sim 2.7 \times 10^{-47} \GeV^4$ for $b_0\sim 5.5\eV$ \cite{ref7}, using the same 
value of the self-coupling parameter as for the $\phi$ field, $\lambda\sim 3\times 10^{-14}$ \cite{ref12,ref13}. 
An attempt is made to obtain
a separate estimate of $b_0$ by coupling the $b$ field to $\nu_\tau$ (with $g_{\nu_\tau}$) \cite{ref7}.
This gives rise to a (presumably largest) neutrino mass $m_{\nu_\tau} = \sqrt{\tilde{m}_{\nu_\tau}^2 +
(g_{\nu_\tau}b_0)^2} \sim \sqrt{(g_{\nu_\tau} b_0)^2} \sim 0.055\eV$, for $b_0 \sim 5.5 \eV$ and
$g_{\nu_\tau} \sim 10^{-2}$ \cite{ref7}, and ``bare'' neutrino mass $\tilde{m}_{\nu_\tau} \sim 0$.
This provides a representation, including the significant role of $\lambda$, of the
often-remarked similarity between the empirical energy scales relevant for a cosmological constant
and for neutrino mass. As illustrated below the effective squared mass of potential $b$ quanta
is negative.
The second assumption here
is that quanta with negative squared mass (i.~e.~superluminal
tachyons \cite{ref14,ref15}) are not present. This effectively prohibits strong long-range forces
due to  exchange of $b$ quanta. 
Thus, we assume that the main effect of a hypothetical coupling to quarks of the
$b$ field is to give a mass contribution to primordial quarks.
A coupling $g$ of the $b$ field to primordial ordinary quarks
gives a quark mass contribution of $gb$; 
subsequently for three confined valence quarks, a nucleon mass contribution of order $3gb$.
We have assumed that primordial quarks have zero bare mass, and we do not consider
possible thermal effects in this paper. We assume that
electroweak symmetry-breaking generates the standard-model MeV mass contribution for light quarks at a later time,
and we assume that this contribution then simply adds to the mass contribution estimated here, $gb$. 
This is the assumption that the electroweak mass term arises from the Higgs vacuum expectation
value times a tiny coupling to the quark field which has acquired a small mass term $gb$.

We consider a very small
perturbation which causes $b$ to change with time.
The essential physical assumption is that $b(t)$, coupled to quarks, moves from a largest value
at $t\sim 0$, toward zero, over cosmological time intervals parameterized by $\ol{t}$,
but comes only  so far as $t\to\infty$, such that the effective squared mass of potential $b$ quanta
approaches zero through negative values. ( The vacuum energy density $\lambda b_0^4$ is assumed
as a cosmological constant.) 
Here, we briefly indicate the consistency of a hypothetical time-dependence for $b(t)$
with an equation of motion, using a particular, hypothetical, generalized force.
An ansatz for the time
dependence is
\be
b^2 = \epsilon b_0^2 + \frac{(1-\epsilon)b_0^2}{(1+(t/\ol{t})^2)^n};\;\;\;
0 <\epsilon < 1,\;\; \dot{b}=\frac{d(b^2)/dt}{2b}
\ee
The time $t$ can be viewed as parameterized by $b(t)$.
\be
\frac{t}{\ol{t}} = \left\{ \left( \frac{(1-\epsilon)b_0^2}{b^2 - \epsilon b_0^2}\right)^{1/n} - 1\right\}^{1/2}
\ee
In the following, we use the Hubble parameter in cosmological time as approximately given by
\be
H(t) = \frac{2}{3t}
\ee
A usual equation of motion (with a term in squared mass $m^2$) is
\be
\ddot b + m^2 b + 3 H \dot b = F = -\frac{dV(b)}{db} = - V'(b)
\ee
$F$ is a generalized force, usually assumed to be derivable from a potential, via $V'(b)$.
With  neglect of $\ddot b$, and with $m^2$ taken as zero, a usual equation of motion
(over relatively brief time intervals in which $H$ changes little), $3 H \dot b = -V'(b)$,
simply represents a variation of $b$ with time, from some initial value at which $V'$ is assumed
to be nonzero, down to a final value at which $V'=0$, presumably a potential minimum, stable in time.
More generally, $F = F(b,\dot b, H(t))$. Then, the analogue of $V''$ is $-F'$, with $\dot b, H(t)$
inserted as explicit functions of $b$. An effective, $b$-dependent (i.~e.~time-dependent) squared
mass in the equation of motion can be viewed as $(m^2 - F')$. With neglect of terms from $\ddot b$,
the equation of motion gives
\be
m^2 = \frac{(-3H\dot b + F)}{b}
\ee
Thus, at a maximum value of $b$ for which $F=0$, $\dot b$ is nonzero because of the term
$m^2$, which can here be viewed as a phenomenological device for causing motion down from a
(metastable) maximum. (Inclusion of terms from $\ddot b$, which must be proportional to
$1/\ol{t}^2$, does not change the essence of the following argument.) Using (1-3)
and (5), for a hypothetical $-F = ( \mu_0^2 b + 3 H\dot{b})$, with $\mu_0^2 = 2n(1-\epsilon)/\ol{t}^2$
fixed by the initial (i.~e.~at $t=0, b=b_0$) condition $F(b=b_0)=0$, one obtains
the following $b$-dependent functions
\ba
-F&=& \mu_0^2 \left[ b - \frac{b_0^2}{b}\left\{ \frac{(b^2 - \epsilon b_0^2)}{(1-\epsilon)b_0^2}\right\}^{\frac{n+1}{n}}\right]\\
-F' & = &\mu_0^2 \left[ 1 + \left(\frac{b_0^2}{b^2}\right)\left\{ \frac{(b^2 - \epsilon b_0^2 )}
{1-\epsilon)b_0^2}\right\}^\frac{n+1}{n} \right. \nn\\
& -  & \left. \frac{2(n+1)b_0^2}{n}\left\{ \frac{(b^2 - \epsilon b_0^2)^{1/n} }{ \left((1-\epsilon)b_0^2\right)^{(n+1)/n}}\right\}\right
]\\
m^2 &=& \mu_0^2 \left[ -1 +2 \frac{b_0^2}{b^2}\left\{ \frac{(b^2 -\epsilon b_0^2)}{(1-\epsilon)b_0^2}\right\}^{\frac{n+1}{n}}\right]
\ea
For illustration, consider $n=1$. Then, the effective, squared mass is
\be
(m^2 - F') = \mu_0^2 \left[ 3 \left(\frac{b_0^2}{b^2}\right)\left\{\frac{(b^2 - \epsilon b_0^2)}{(1-\epsilon)b_0^2}\right\}^2
- \frac{4 (b^2 - \epsilon b_0^2)}{(1-\epsilon)^2 b_0^2}\right]
\ee
This quantity is negative at $b^2=b_0^2$ (at $t=0$), $\mu_0^2 \left[ 3- (4/(1-\epsilon))\right] = -(2/\ol{t}^2) (1+3\epsilon)$.
It goes to zero through negative values, as $b^2 \to \epsilon b_0^2$ (as $t\to \infty$).

With eq.~(1), we have for time-dependent, approximate contributions to $m_p$.
\ba
{\mathrm{at}}\;\; t \sim 0  && 3gb_0\nn\\
{\mathrm{at}}\;\; t\sim \overline{t} && 3g\sqrt{(1+\epsilon)/2}\; b_0 \\
{\mathrm{at}}\;\; t\to \infty &&  3g \sqrt{\epsilon} b_0\nn
\ea
The direction of mass change is downward in the model.
There is a definite decrease as $t\to \infty$, the magnitude 
is related to $b_0$.
Even  with a sizable effective
``magnification'' factor \footnote{
With reference to possible ``magnification'' of the effective $g$, it might be useful
to note that the confinement of quarks at $\sim 10^{-6}\secs$, does involve electroweak
mass $\sim \mbox{MeV}$, being substantially increased to constituent quark mass. The QCD
energy-scale parameter is $\Lambda \sim 220 \MeV$.
}, $g\sim 10^3$, one obtains a small scale of
mass change, $\sim \mbox{keV}$. Conceptually, this is related through $b_0$ to a  very small
cosmological constant, and possibly to a very small neutrino mass.
The electron mass can change, but the leptonic $b$ coupling
may be like that estimated for neutrinos, $g_l \sim 10^{-2}$. Thus, the hypothetical
downward change in $(m_p/m_e)$ is probably controlled by the downward change in $m_p$.
The parameter $\overline{t}$ might be related to other dynamical quantities in the
early universe. It can be connected with the ratio of vacuum expectation values,
which ratio is numerically closely given in terms of 
the very small parameter $\lambda$ \cite{ref11},
that scales the primordial, vacuum energy densities: $\lambda^2 \sim b_0/\phi_0
\sim 5.5\eV/10^{18}\GeV \sim 5.5\times 10^{-27}$.
(Here $\phi_0 \sim 10^{18}\GeV$,
at an assumed minimum of zero for the inflaton effective potential.) 
Numerically, $\overline{t} \sim 3 \times 10^{16}\secs \sim
(10^{-36}\secs)\times(1/\lambda^4)$, where $10^{-36}\secs$
is the time near the end of inflation generated by the $\phi$ field \cite{ref11} \footnote{
As counted from the time of $\phi$ leaving its value $\simgt M_{\mathrm{Pl}}$ at the effective potential maximum.
}.
With this representation for $\ol{t}$, the expansion scale factor is given approximately as 
$a(\overline{t}) \sim (\overline{t}/10^{-36}\secs)^{1/2} \sim (1/\lambda^2)$ \footnote{It is interesting to note that the
necessary minimal value of the expansion scale factor for an initial inflation over
$\Delta t$ is a similar number. That is $a_{\mathrm{infl}} (\Delta t) = e^{H_{\mathrm{in}} \Delta t}  \sim
1/\lambda^2 \sim 2 \times 10^{26}$, for $\Delta t \sim 1/H_{\mathrm{in}} \times \ln (1/\lambda^2)$,
where $H_{\mathrm{in}}$ is the Hubble parameter as fixed by the initial vacuum energy
density of the inflaton field, which is proportional to $\lambda$.
}.

To recapitulate, the
unusual assumption is that the $b$ field can move from a  maximum value 
toward the value zero, over cosmological time intervals, but comes only
a part of the way as $t\to \infty$. 
The second essential assumption is that 
quanta
of the $b$ field with negative squared mass are not present to induce strong long-range
forces. In the numerical estimates, an essential number is the empirically very small
self-coupling parameter $\lambda$ for the inflaton $(\phi)$ field; this parameter is common
to the $b$ field self coupling in the model.

We conclude with the remark that a general idea seems to receive support from the
possible small decrease of $(m_p/m_e)$ with cosmological time \cite{ref1}. 
This is that there is a small energy scale $b_0$, associated with the early
universe \cite{ref7,ref11}, in addition to the usual very high energy scales,
i.~e.~inflaton mass and radiation temperature. Effects of the $b$ field and
of the $\phi$ field are related, when depending upon the single parameter $\lambda$ \cite{ref11}.
The above smallness of $\lambda$ from the ratio of vacuum expectation values, can give
the relatively slow evolution of the field and energy density on a short time
scale for $\phi$, and possibly on a long time scale for $b$ from a relationship
of the very small parameter $\lambda$ to a cosmological time parameter $\ol{t}$. Clearly,
if $\lambda$ were to approach zero, then the  vacuum energy density in the $b$ field
would approach zero, and  $\ol{t}$ would approach infinity.
The small energy scale is capable of relating a very small  cosmological
constant today
and a small decrease of mass over cosmological time. 

\section*{Appendix I}

At the end of inflation, there are two very high, energy-scale parameters, $\phi_0\cong 10^{18} \GeV$
and the initial radiation temperature $T_0 \cong 10^{15.5} \GeV$. But there is also a
very small, dimensionless parameter $\lambda$, of the order of $10^{-14}$. \footnote{Heuristically,
this can be seen by equating the maximum energy density in the inflaton field, to the
initial (maximum) radiation energy density, which appears just after inflation.
Then, $\sim \lambda M_{\mathrm{Pl}}^4 \sim T_0^4$ gives $\lambda \sim 10^{62}\GeV^4/10^{76}\GeV^4 =
10^{-14}$. }
A hypothetical small energy scale $ b_0 \cong 5.5 \eV$ is obtained from a relation
$\lambda^2 = b_0/\phi_0 = 5.5 \times 10^{-27}$ ( for $\lambda = 7.4\times 10^{-14}$).
There are then two time-scale parameters: $1/\sqrt{\lambda} \phi_0 = 2.4 \times 10^{-36} \secs$
at the end of inflation; and $1/\sqrt{\lambda} b_0 = 4.4 \times 10^{-10} \secs$. Over
the latter time scale, the radiation energy scale evolves downwards to a relatively
low scale
\be
T_0\left( \frac{2.4 \times 10^{-36}\secs}{4.4\times 10^{-10}\secs }\right)^{1/2} =
h = 234 \GeV
\ee
This scale is almost the empirical Higgs energy scale (the assumed nonzero vacuum
expectation value v) of the standard model. The coupling $g_e$, of $v/\sqrt{2}$ 
to electrons, is empirically another small number: $g_e^2 \cong 9\times 10^{-12}$.
Again, a ratio of energy scales almost gives this very small number $g^2 = b_0 /h 
= 23.5 \times 10^{-12}$.\footnote{Another ratio of energy scales with $b_0$ possibly
gives a dimensionless measure of electromagnetic interaction strength, that is
$\alpha^2 = 5.3\times 10^{-5} \cong b_0/m_e \cong 5.5 \eV/0.5\MeV = 1.1 \times 10^{-5}$;
$ \alpha^2 \propto\sqrt{\lambda}$. The ratio of a nuclear (QCD) energy scale to
an atomic (binding) energy scale is $1/\sqrt{\lambda}$, giving a nuclear scale
of order 40 MeV for an atomic scale of order 10 eV, with $\lambda\cong 7\times 10^{-14}$
(clearly, $\lambda$ must be much less than unity).}
In summary, the above emphasis on the role of $\lambda$ allows for simply 
\ba
b_0 &=& \lambda^2 \phi_0 \propto\lambda^2 \nn\\
h &=& \lambda \phi_0 (T_0/\phi_0) \propto\lambda\\
g^2 &=& \lambda (\phi_0/T_0) \propto \lambda ,\;\;\mbox{so}\;\;g \propto \sqrt{\lambda}\nn
\ea
The essential parameters are $\phi_0, T_0$ and $\lambda^2 $ ( or $b_0$). Here, it is
the extreme smallness of $\lambda$ which relates the high energy scale $\phi_0$ ( or $T_0$)
to the much lower energy scale $h$. Also, the smallness of a minimal Higgs coupling
is related to the smallness of $\sqrt{\lambda}$.\footnote{The contribution of zero-point energies in massless
quanta to the vacuum energy density, may be initially limited by a specific large
dimension of the inflation region, $(1/\sqrt{\lambda}\phi_0) \times (1/\lambda^2)
= 1/\sqrt{\lambda} b_0$, that is the very rapid expansion may result in a
density of order $(\sqrt{\lambda} b_0)^4$. (The empirical, apparent cosmological
constant is $\sim \lambda b_0^4$.)}

\section*{Appendix II - ''Cosmic coincidence''}
The present ratio of vacuum energy density $\rho_\Lambda$, to approximate dark-matter energy
density $\rho_{\mathrm{d.m.}}$, empirically $\rho_\Lambda/\rho_{\mathrm{d.m.}}\sim 3$, appears to
be a curious accident, in particular when these densities are assumed to originate in 
completly different dynamics, and the matter density falls with the expansion of the universe.
Here, we note that the ratio of order unity is not particularly accidental, when the two densities
are considered together as functions of only two quantities: the very large energy scale
$\phi_0$, and the very small dimensionless parameter $\lambda$, using relations in the text.
We have \footnote{For $\lambda  \sim 7.4\times 10^{-14}$, Eq.~(13) gives $\rho_\Lambda$ above the
empirical value by only a factor of about 2.5.} 
\ba
\rho_\Lambda &=& \lambda b_0^4 = \lambda (\lambda^2 \phi_0)^4 = \lambda^9 \phi_0 \\
\rho_{\mathrm{d.m.}} &=& (\rho_{\mathrm{d.m.}})_0 \left\{\left(\frac{10^{-36}\secs}{\ol{t}}\right)^{3/2}
\left(\frac{\ol{t}}{10^{11}\secs}\right)^{3/2}\left(\frac{10^{11}\secs}{4\times 10^{17}\secs}\right)^2\right\}\nn\\
&=& \left( \frac{1}{2}f(\lambda)m_\phi^4 \right) \left\{ (\lambda^4 )^{3/2} \times(\sim \lambda^{0.39})\right\}\\
&=& (\lambda^{2.5} \phi_0^4) \left\{ (\lambda^6)\times (\sim \lambda^{0.39})\right\}\nn\\
&\cong& \lambda^{8.89} \phi_0^4\nn
\ea
In (14), $(\rho_{\mathrm{d.m.}})_0$ is the initial energy density for inflatons created at $\sim 10^{-36}\secs$.
This density is written in terms of the inflaton mass $m_\phi \cong \sqrt{\lambda} \phi_0$, as 
$ m_\phi^2/2 \times (f(\lambda)m_\phi^2)$ with an assumption for the fluctuation scale \cite{ref8},
$\lambda < f(\lambda) < 1$. \footnote{ This is consistent with an estimate of the size of $f$ from
production of massive dark matter by a time-varying gravitational field, when $H$ is of order
$m_\phi$.\cite{ref11}} As a definite example, we used in (14) the geometric mean
$f(\lambda) =\sqrt{\lambda}$. The terms bracketed as $\{\ldots\}$ give the time evolution to the
present age of the universe, $\sim 4\times 10^{17}\secs$. The first term explicitly uses
the text's hypothetical $\lambda$ dependence of $\ol{t}$, for evolution to the time $\ol{t} =
3\times10^{16}\secs$, which is about a characteristic time for galaxy-halo formation from dark
matter. The next two terms give the evolution from $\ol{t}$ to the present, with
$\sim 10^{11} \secs$ taken as an approximate time of matter dominance. These two factors
are expressed as an effective, small power of $\lambda$. Eqs.~(13,14) give$^{F8}$ 
\be
\rho_\Lambda/\rho_{\mathrm{d.m.}} \sim 1/30
\ee
This differs from the approximate empirical value by only a factor of about 90. From this point
of view, a ratio of order of unity may be explainable in terms of only two primary, dynamical
quantities, $\phi_0$ and $\lambda$ (or $b_0$).

\end{document}